\documentclass[aps,prd,nofootinbib,twocolumn,floatfix,letterpaper,superscriptaddress,showpacs]{revtex4}
\usepackage{graphicx}\usepackage{natbib}\usepackage{amsmath}
\pdfoutput=1
\usepackage{times}

\newcommand{\mnras}{Mon.~Not.~Roy.~Astron.~Soc.\ }
\newcommand{\aap}{Astron.~Astrophys.\ }

\begin{document}

\title{On TeV Gamma Rays and the Search for Galactic Neutrinos}

\author{Matthew D. Kistler}
\affiliation{Kavli Institute for Particle Astrophysics and Cosmology, Stanford University, SLAC National Accelerator Laboratory, Menlo Park, CA 94025}

\date{November 16, 2015}

\begin{abstract}
The IceCube neutrino discovery presents an opportunity to answer long-standing questions in high-energy astrophysics.  For their own sake and relations to other processes, it is important to understand neutrinos arising from the Milky Way, which should have an accompanying flux of gamma rays.  Examining {\it Fermi} TeV data, and applying other constraints up to $>\,$1~PeV, it appears implausible that the Galactic fraction of the IceCube flux is large, though could be present at some level.  We address Sgr~A*, where the TeV--PeV neutrinos may outrun gamma rays due to $\gamma \gamma$ opacity, and further implications, including dark matter and cosmic-ray electrons.
\end{abstract}

\maketitle

\section{Introduction}
Cosmic-ray interactions in the Milky Way produce mesons decaying to secondary particles (e.g., \cite{Feenberg1948,Fermi1949,Richtmyer1949}) forming a flux of neutrinos that has long been of great interest to detect (e.g., \cite{Greisen:1960wc,Reines:1960we,Markov1961,Stecker:1978ah,Roberts:1992re,Barwick:1991ur,Gaisser:1994yf,Learned:2000sw,Halzen:2002pg,Ahrens:2002dv,Katz:2006wv,Desai:2007ra,Adrian-Martinez:2015ver}).  The recent discovery by IceCube of high-energy astrophysical neutrinos \cite{Aartsen2013,Aartsen2013b,Aartsen2014,Kopper2015,Aartsen2015,Niederhausen2015,Aartsen2015b,Radel2015} implies a particularly large flux of $\sim\! 10 \!-\! 100$~TeV neutrinos \cite{Aartsen2015,Niederhausen2015}.  Naturally, establishing the fraction that arises from our Galaxy will benefit both studies of high-energy processes within the Milky Way (e.g., \cite{Gaisser:1990vg, Alvarez-Muniz:2002tn, Albuquerque:2001jh, Costantini:2004ap, Bednarek:2004ky,Crocker2005, Candia:2005nw,Lipari:2006uw,Kistler2006,Horns2006,Prodanovic:2006bq,Kappes:2006fg,Beacom:2007yu,Evoli:2007iy,Halzen:2008zj,Villante:2008qg,Yamazaki:2008fj,GonzalezGarcia:2009jc,Ahlers:2009ae,Anchordoqui:2009nf,Yuan:2010vi,Tchernin:2013wfa,Mandelartz:2013ht,Fox:2013oza,Gupta:2013xfa,Ahlers2014,Taylor:2014hya,Kachelriess:2014oma,Neronov:2014uma,Gaggero:2015xza,Ahlers:2015moa,Neronov:2015osa}) and determinations of the extragalactic component to address associated problems (e.g., \cite{Kistler2014,Anchordoqui2014,Baerwald2015}).

We discuss several constraints on the gamma rays that generally will accompany these neutrinos in relation to the IceCube data.  For instance, we see that the IceCube flux at $\sim\!10$~TeV is now rather near to electromagnetic shower data from electron searches by the air-Cherenkov telescopes (ACTs) HESS \cite{Aharonian2008,Aharonian2009b} and VERITAS \cite{Staszak2015}.  At higher energies, constraints are derived from air shower arrays.

Though of smaller proportion than ACTs, and much less tonnage than IceCube, the {\it Fermi} Large Area Telescope (LAT) \cite{Atwood:2009ez} can partially compensate with a long observing time.  Of great benefit to diffuse studies is the instrumental veto power against cosmic rays and a large field of view with fairly uniform exposure over the sky to $E_\gamma \!>\! 1$~TeV with Pass~8 \cite{Atwood:2013rka}.

For rough comparison, the IceCube neutrino effective area is $A_\nu^{\rm eff} \!\sim\! 2~{\rm m}^2$ at $E_\nu \!\sim\! 10$~TeV for the all-sky search reporting 172 shower events in \cite{Niederhausen2015}, comparable to the LAT geometric area.  Even without extrapolating a gamma-ray flux back much further to lower energies or accounting for $A_\nu^{\rm eff}$ increasing with $E_\nu$, this suggests that there should be no shortage of gamma rays incident upon the LAT if there is an appreciable Galactic contribution to the IceCube flux.

We examine {\it Fermi} LAT data to assess the current picture of the entire TeV sky, showing emission highly-concentrated along Galactic plane and within $\lesssim\! 30^\circ$ of the Galactic Center (GC).  We do so making use of all likely TeV events, even though leptonic TeV sources are plentiful in the Milky Way, and often extended, to avoid subtracting off any possible neutrino-associated gamma rays (cf., e.g., \cite{Neronov:2014uma}).  Since there should be little attenuation due to photon backgrounds \cite{Porter2006}, this constrains the amount of Galactic neutrino production.

We derive upper limits from various regions and further break down the IceCube events into maps by estimated neutrino energy for comparison.  We also discuss several other possibilities, such as a Galactic proton spectrum harder than locally measured, dark matter, and constraints on contributions of residual TeV gamma rays to ACT $e^- \!+\! e^+$ data.

One possible exception to such limits is the supermassive black hole complex at the GC, in which gamma rays would be strongly attenuated within the inner accretion flow of Sgr~A*, while neutrinos can escape to very high energies.  We examine the possibility of producing neutrinos up to PeV energies, as may be hinted at by IceCube data \cite{Aartsen2013,Aartsen2013b}, and the gamma-ray flux in relation to the TeV signal seen from the GC by HESS \cite{Aharonian2009,Viana2013} and VERITAS \cite{Archer2014,Smith2015}.

\begin{figure*}[t!]
\vspace*{-0.85cm}
\includegraphics[width=1.9 \columnwidth,clip=true]{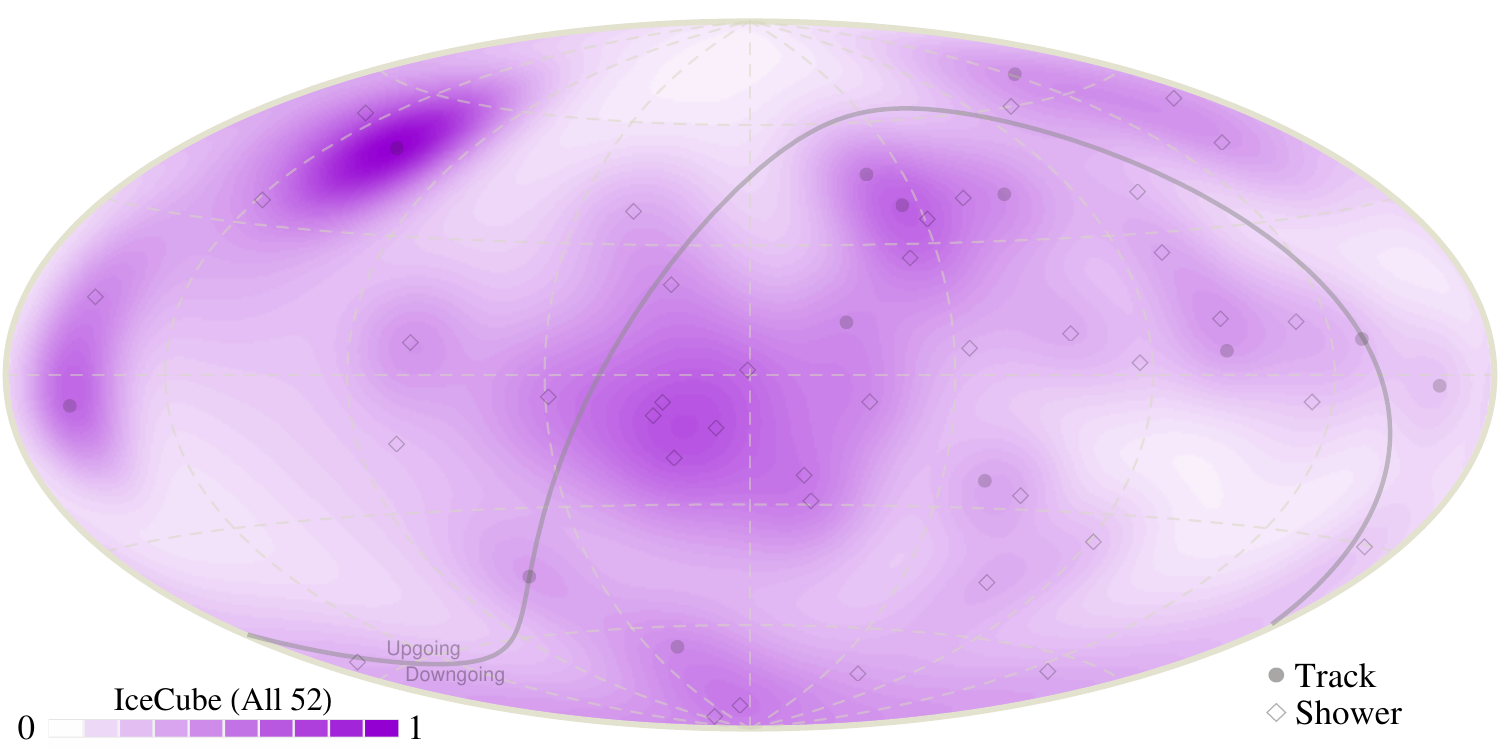}\\
\vspace*{-0.2cm}
\caption{Arrival directions of 52 neutrino events detected by IceCube \cite{Aartsen2013b,Aartsen2014,Kopper2015} (in Galactic coordinates with horizon demarcating upgoing and downgoing directions).  To obtain an estimated density distribution ({\it contours}), we replace each event with a Gaussian (showers: median angle, muons: $10^\circ$ for display) and weighting by a relative exposure corresponding to Earth opacity (see text).
\label{iceall}}
\end{figure*}

\section{Once Upon a Time in Antartica}
We begin with the most recent compilation of available IceCube contained neutrino events \cite{Aartsen2013b,Aartsen2014,Kopper2015}, which give a better idea of the neutrino energy than throughgoing muons, in order to construct an approximate density distribution (as in \cite{Yuksel2012}).  Lacking event-by-event angular distributions, we use a Gaussian with the median angular uncertainty \cite{Aartsen2013b,Aartsen2014,Kopper2015} as the width for shower events.  Obviously, for a likelihood analysis one would want to make use of the precise angular data; however, the hierarchy between tracks ($\sim\! 1^\circ$) and showers ($\gtrsim\! 10^\circ$) presents a difficulty for display.  To be closer to the same scale, we use $10^\circ$ for tracks (which might have been showers anyway if not for neutrino oscillations).

In Fig.~\ref{iceall}, we show the density of 52 events (not including events 28 and 32 as possible background; see \cite{Kopper2015} for a recent skymap showing shower and track numbers).  Here, we have attempted to compensate for loss of neutrinos due to Earth opacity by reweighting each event based on the neutrino energy via $\sigma_{\nu N}(E_\nu)$ \cite{Gandhi:1995tf,Gandhi:1998ri} and column of Earth traversed \cite{Dziewonski:1981xy} to determine $e^{-\tau_\oplus}$.  Now, we do not know the actual neutrino energy, of course, which is needed in determining $\sigma_{\nu N}$.

For showers, we take $E_\nu \!=\! E_{\rm em}$, although this will be an underestimate for the unknown fraction of events due to $\nu_\tau$ or neutral-current interactions, which would have a higher true $E_\nu$.  For tracks, we assume $E_\nu \!=\! 3\, E_{\rm em}$ to allow for the fractional energy loss of the muon prior to exiting the detector volume, although this may contain significant dispersion \cite{Aartsen:2013vja}.  While imperfect, this compromise appears preferred to, e.g., adding a proportionate number of fake events or increasing the Gaussian width in the upgoing region.  Later, we will split these into energy ranges motivated by classes of gamma-ray data.  We note that the number of events is larger in the downgoing region of the sky (as denoted), where there is less Earth attenuation and downgoing muons can form a residual background, and that no obvious structure is evident at this level.

\begin{figure*}[t!]
\vspace*{-0.85cm}
\includegraphics[width=1.9 \columnwidth,clip=true]{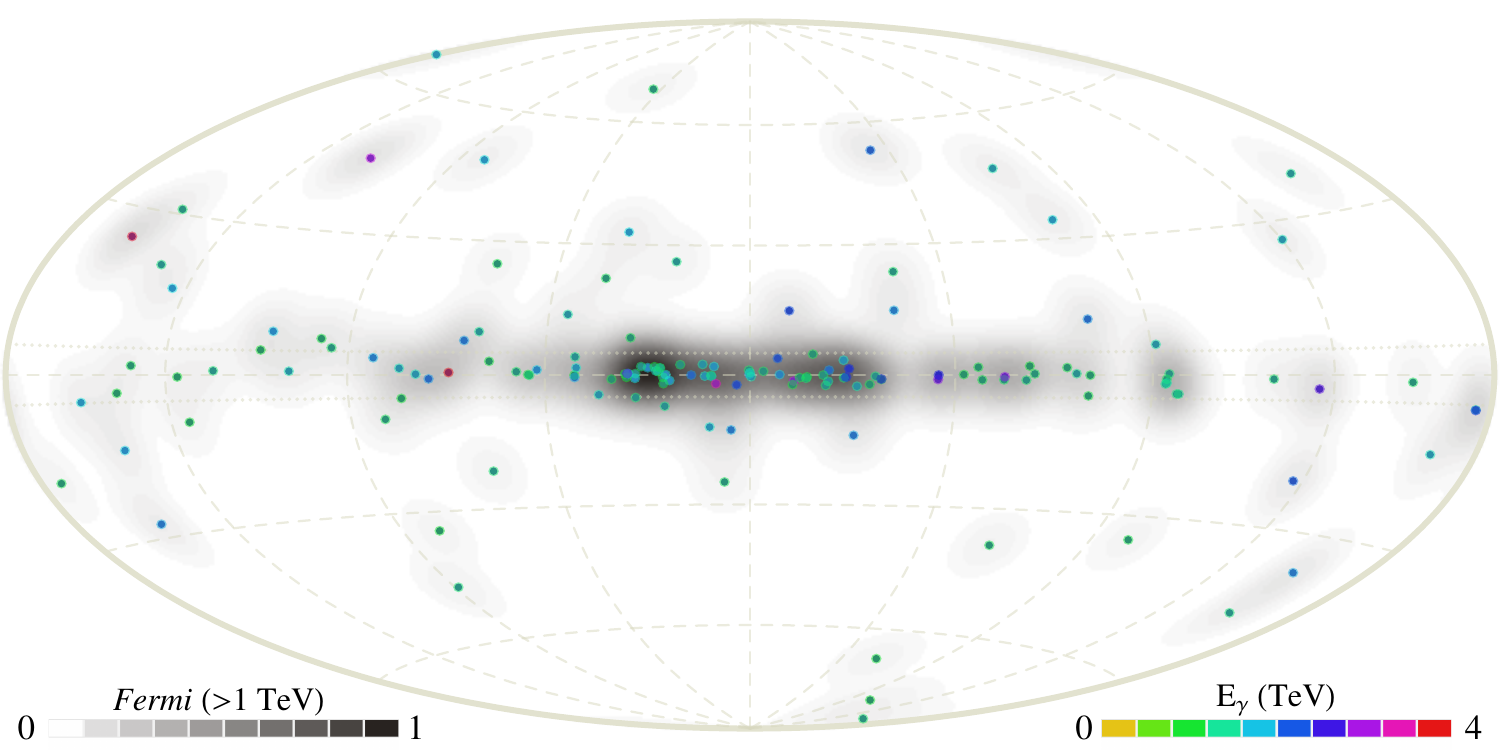}
\vspace*{-0.2cm}
\caption{{\it Top:} Gamma rays with $E_\gamma \!>\! 1$~TeV detected by {\it Fermi} LAT (Pass~8 data; colored by $E_\gamma$).  While the LAT data have a PSF of $\lesssim\!0.2^\circ$, to obtain a density distribution closer to the IceCube map we replace each event with a $5^\circ$ Gaussian weighted by the LAT exposure (note though that IceCube muon tracks typically have angular resolution comparable to LAT gamma rays) with the dotted line denoting $b \!=\! \pm 5^\circ$.
\label{fermitev}}
\end{figure*}

\section{A {\it Fermi} Full of Gamma Rays}
We use {\it Fermi} LAT Pass~8 data, which has greatly enhanced performance at high energies to $E_\gamma \!>\! 1$~TeV \cite{Atwood:2013rka,LATpage}, from October 30, 2008 -- October 15, 2015.  For our purpose, we use the class with the cleanest gamma-ray event sample (P8R2\_ULTRACLEANVETO\_V6) with recommended cuts to avoid the Earth limb (zenith angle $< 90^\circ$) from the entire sky.  We limit use to energy dispersion sub-classes EDISP1, EDISP2, and EDISP3 (excluding the lowest class EDISP0).

Fig.~\ref{fermitev} shows the positions of 153 events with $E_\gamma \!>\! 1$~TeV in this sample, with 97 within $|b| \!<\! 5^\circ$ of the Galactic plane, 112 within $|b| \!<\! 10^\circ$, and 123 within $|b| \!<\! 20^\circ$.  In forming a density for comparison with the IceCube maps, using the track event $\sim\! 1^\circ$ scale results in a map looking much like the points, while using the $\sim\! 10^\circ$ of showers blurs even coarse features.  With $\sim\! 2/3$ of events falling within $5^\circ$ of the Galactic plane, we use $5^\circ$ as a compromise.  Here, we reweight each event based on the LAT exposure, which is decreasing with energy above 1~TeV.  We create exposure maps at 500~GeV intervals and linearly interpolate at intermediate energies.

Our interest is in comparing to the neutrino flux, for which we do not know the origin, so we aim for inclusiveness and remove no foreground emission and subtract no Galactic source regions, since these could all potentially be related to the IceCube signal.  While extragalactic sources should be highly attenuated, for completeness, we subtract off three events within $0.2^\circ$ of extragalactic Second Hard Fermi-LAT (2FHL) catalog \cite{Ackermann:2015uya} sources (Mrk~421, IC~310, and 1ES 0347--121) which are thus likely not Galactic.  We also consider only photons with $E_\gamma \!<\! 3.16$~TeV, removing three more to leave 147 in total.

Since we lack definitive information about the nature of possible sources, we use these data to derive a $2\sigma$ limit on the all-sky TeV flux, shown in Fig.~\ref{SED}.  Here we have allowed for 20\% systematic uncertainty in the exposure (and different variations in counts would not change much).  Considering only $|b| \!>\! 20^\circ$, we are left with 27 photons with the commensurately larger statistical uncertainty included in Fig.~\ref{SED}.

We also check regions with diffuse measurements reported by ground arrays, which have less direct means of accounting for cosmic-ray backgrounds.  We compare to Milagro at $|b| \!<\! 5^\circ$, $40^\circ \!<\! l \!<\! 100^\circ$ at $\sim\! 3.5$~TeV \cite{Atkins:2005wu}, which also has a higher-energy measurement \cite{Abdo:2008if}, and ARGO-YBJ \cite{Bartoli:2015era}.  These appear roughly consistent with a soft spectrum back to 1~TeV, although we have not attempted to mask source regions and LAT TeV statistics are limited in the outer plane.

\section{For a Few Neutrinos More}
\begin{figure*}[t!]\vspace*{-0.2cm}
\includegraphics[width=1.9\columnwidth,clip=true]{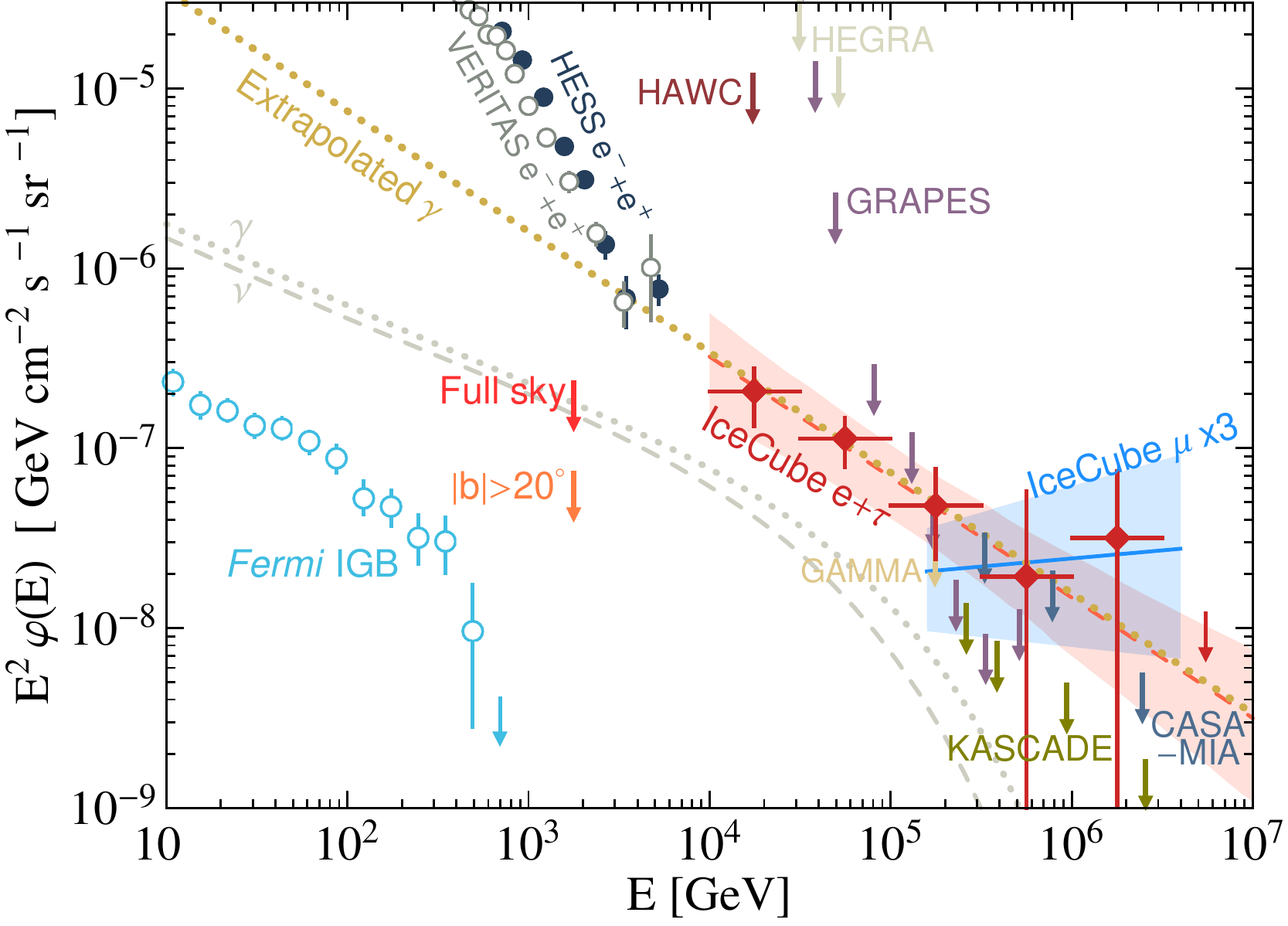}
\vspace*{-0.35cm}
\caption{IceCube neutrino fluxes based on $\nu_e + \nu_\tau$ (\cite{Niederhausen2015}; {\it diamonds}, {\it dashed line + band}) and $\nu_\mu$ (\cite{Radel2015}; {\it solid line + band}).
We show the {\it Fermi} IGB \cite{Ackermann2015} and our limits on isotropic and $|b| \!>\! 20^\circ$ $E_\gamma \!>\! 1$~TeV {\it Fermi} fluxes.
Limits at $\gtrsim\! 10$~TeV are on isotropic gamma-ray fluxes from HAWC \cite{Pretz2015}, HEGRA \cite{Aharonian2002}, GRAPES-3 \cite{Gupta}, GAMMA \cite{Martirosov2009}, KASCADE \cite{Schatz2003}, and CASA-MIA \cite{Chantell1997}.  
We also include $e^- + e^+ +(\gamma)$ spectra from electromagnetic air showers from HESS~\cite{Aharonian2008,Aharonian2009b} and VERITAS~\cite{Staszak2015}.
To these we compare saturating $p p$ neutrino and gamma-ray fluxes from $dN/dE_p \!=\! E_p^{-2.5}e^{-E_p/1~{\rm PeV}}$ ({\it labeled}) and a gamma-ray flux extrapolated from the IceCube fit assuming $\pi^+$ $\pi^-$ $\pi^0$ equality ({\it dotted}).
\label{SED}}
\end{figure*}

We also show in Fig.~\ref{SED} the most recently presented IceCube spectral data.  These include the aforementioned $\nu_e + \nu_\tau$ shower search that uncovered more $\gtrsim\! 10$~TeV events \cite{Niederhausen2015}.  Muon data beginning at higher energies imply a harder spectrum for $\nu_\mu$ \cite{Aartsen2015b}, $\sim\! E_\nu^{-1.9}$, shown from \cite{Radel2015}.  We have assumed equality between flavors and $\nu/\bar{\nu}$ in scaling the data in Fig.~\ref{SED}.  Note that these these fluxes do not directly correspond to the events used in Fig.~\ref{iceall}, though.

First, a simple extrapolation of the soft $\sim\! E_\nu^{-2.67}$ spectral fit \cite{Niederhausen2015} to gamma rays assuming assuming equal numbers of $\pi^+$, $\pi^-$, and $\pi^0$ \cite{Kistler2006} clearly overshoots our TeV limits.  At the high-energy end, we include the collection of limits from \cite{Kistler2009} on isotropic gamma-ray fluxes up to $\sim\! 10^7\,$GeV \cite{Aharonian2002,Chantell1997,Schatz2003,Gupta}, adding to these results from HAWC \cite{Pretz2015} and GAMMA \cite{Martirosov2009} (see also \cite{Lacki2010,Ahlers2014}), some of which cut below this flux.

In between these data is where the present neutrino flux is largest.  Since the properties of air showers induced by electrons and gamma rays are very similar, we also include results from HESS \cite{Aharonian2008} and VERITAS \cite{Staszak2015} measurements of showers in search of high-energy electrons.  The sight-lines used tend to be from regions around targets off the Galactic plane to avoid gamma-ray emission and it is assumed that the extragalactic TeV gamma-ray flux is attenuated \cite{Aharonian2008}, as implied by the {\it Fermi} isotropic gamma-ray background (IGB) \cite{Ackermann2015}.

We see these drop sharply to $\sim\!10$~TeV and can be considered as upper limits on the flux of gamma rays in this range (as in \cite{Kistler2010}), already approaching the IceCube flux level.  Further, HESS claims that the gamma-ray fraction is likely $\lesssim\! 10$\% (systematic uncertainties could reach as high as $\sim\! 50$\%) \cite{Aharonian2008}, which would press down greatly even on spectra that are much harder or at a lower flux level in the $\lesssim\! 10$~TeV range.  It seems fair to conclude that any appreciable Galactic contribution to IceCube up to $\sim\! 10$~TeV should also be confined to the plane.

We construct neutrino and gamma-ray spectra based on $pp$ scattering \cite{Kelner2006} assuming an $\sim\! E_p^{2.5}$) proton spectrum with an exponential cutoff at $1$~PeV scaled to just below our all-sky TeV limit for comparison.  This is harder than the proton spectrum measured at Earth, though with a similar cutoff energy.  Even so, it falls well below the IceCube flux.

\begin{figure*}[t!]
\vspace*{-0.85cm}
\includegraphics[width=1.8 \columnwidth,clip=true]{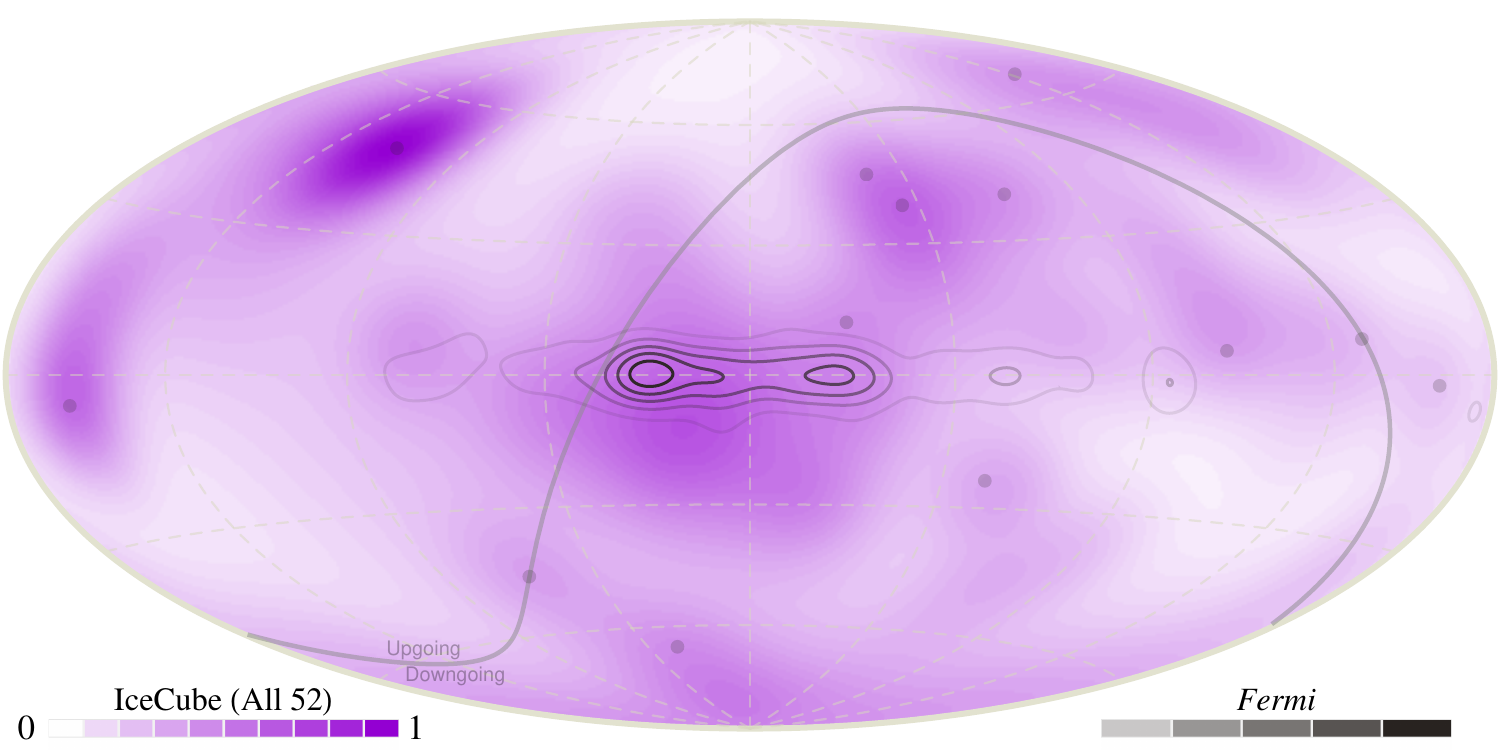}\\
\vspace*{-0.1cm}
\includegraphics[width=1.8 \columnwidth,clip=true]{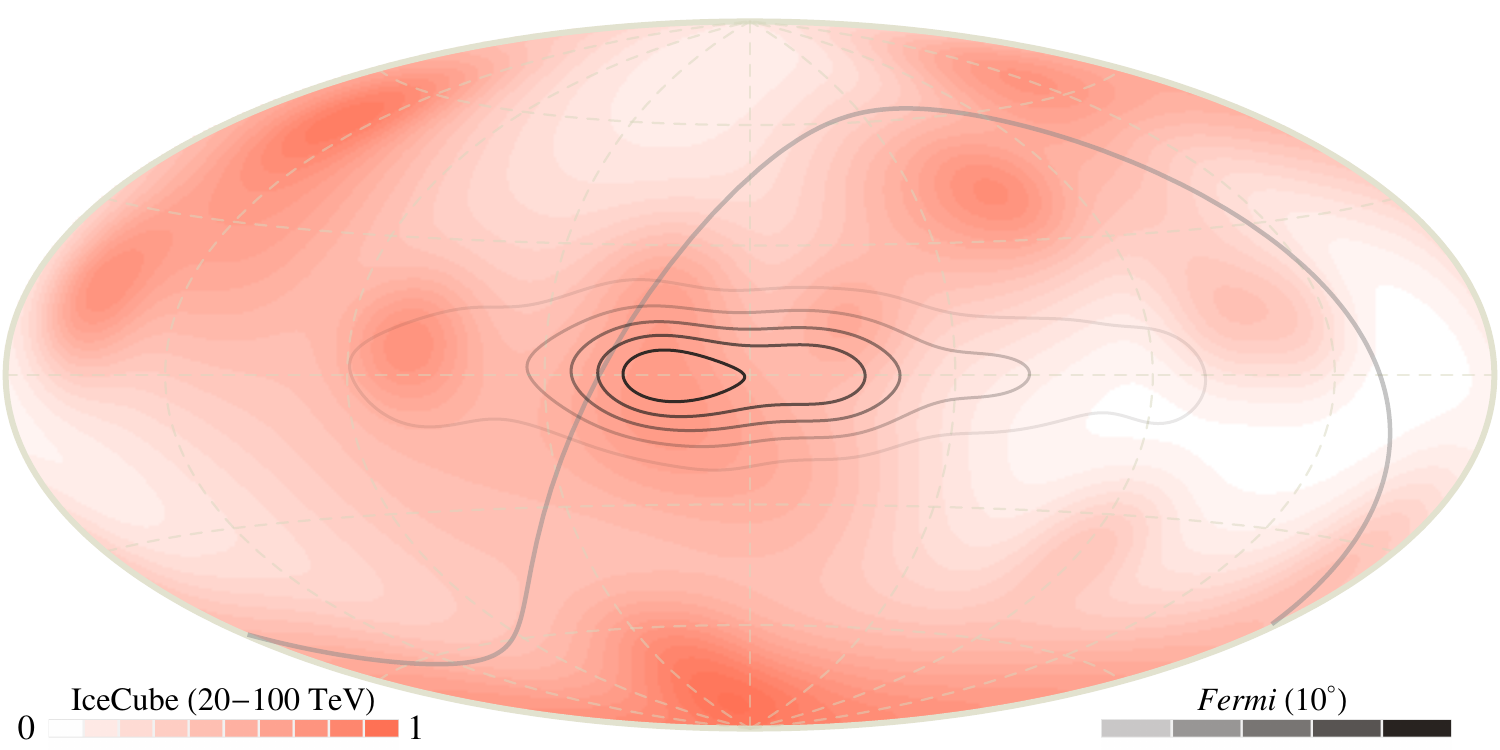}\\
\vspace*{-0.1cm}
\includegraphics[width=1.8 \columnwidth,clip=true]{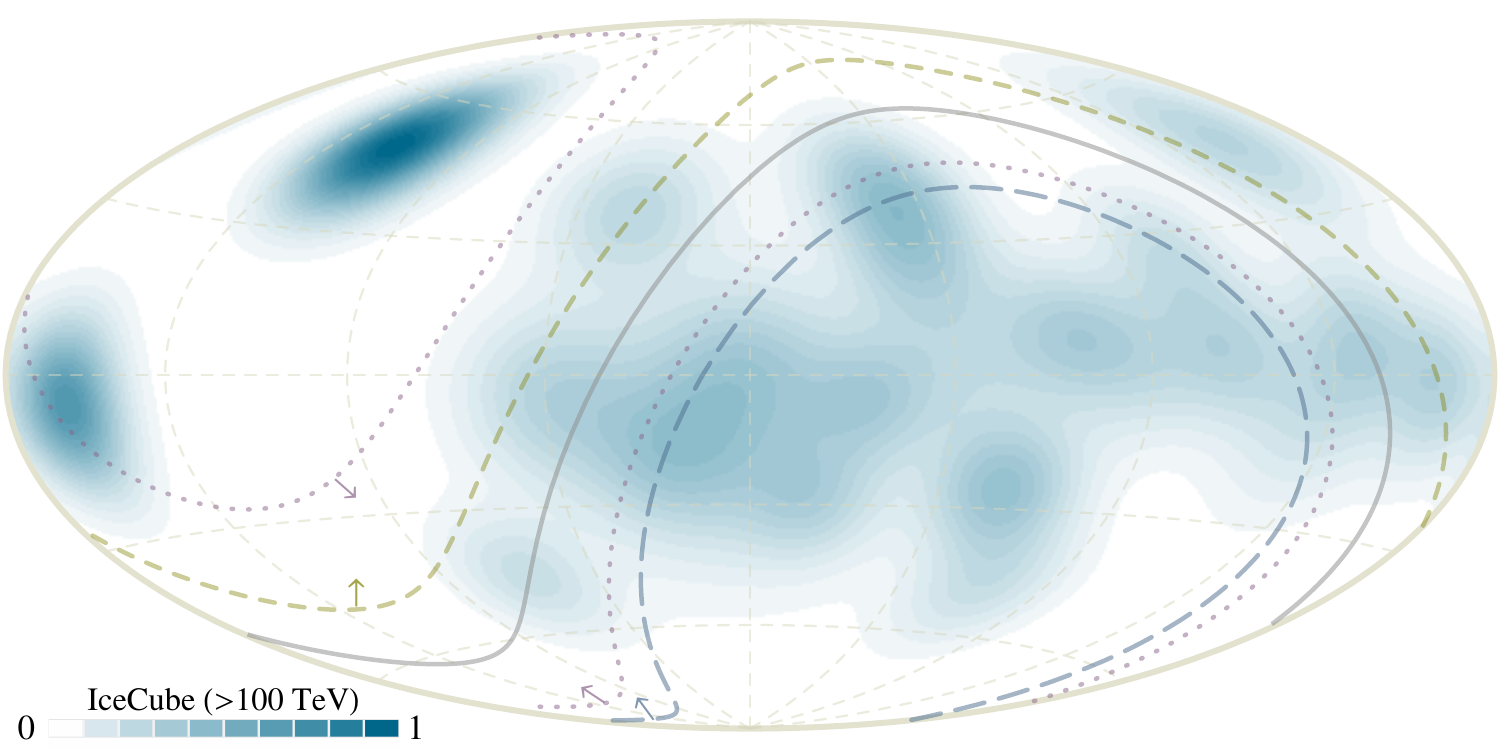}
\vspace*{-0.2cm}
\caption{{\it Top:} Contours of {\it Fermi} TeV density in Fig.~\ref{fermitev} on the density distribution of 52 IceCube events from Fig.~\ref{iceall} (with track positions; {\it dots}).
{\it Middle:}  IceCube density for 27 events with estimated $E_\nu \!=\! 20$--100~TeV, with {\it Fermi} TeV density contours using a $10^\circ$ Gaussian (closer to the resolution of the 23 showers in this set).
{\it Bottom:} IceCube $>\! 100$~TeV event density, with sky coverages indicated for CASA-MIA (\cite{Chantell1997}; above {\it long-dashed blue}), KASCADE (\cite{Schatz2003}; above {\it short-dashed green}), and GRAPES (\cite{Gupta}; between {\it dotted purple}) gamma-ray limits from Fig.~\ref{SED}.
\label{icetev}}
\end{figure*}

In the top panel of Fig.~\ref{icetev}, we plot contours of the Fig.~\ref{fermitev} TeV density on the density from Fig.~\ref{iceall} along with IceCube track positions.  We see no track events within the median angular error of Galactic plane.  The region within the contours contains $\sim\! 2/3$ of the total TeV photons and zero tracks.

Additionally, sources have been detected by many TeV gamma-ray experiments.  Milagro reported $\gtrsim\!10$~TeV sources that tend to be coincident with {\it Fermi} pulsars \cite{Abdo:2009ku}, suggesting leptonic pulsar wind nebula (PWN) emission.  This also supports the idea that many unidentified TeV sources are PWNe (e.g., \cite{Yuksel2009,Tibolla:2012in,Kargaltsev:2013gaa,Acero:2013xta}), especially if {\it Fermi} only sees a fraction of pulsars due to beaming of the gamma-ray emission.

Even in our limited sample, we find events likely associated with the Crab and Vela~X PWNe and supernova remnants that may be leptonic in the TeV (e.g., RX J1713.7--3946 and Vela Jr.; \cite{Acero:2015caa}).  In Figs.~\ref{fermitev} and \ref{icetev}, the most prominent clump ($\sim\! 20^\circ$ to the left of the GC) is likely associated with the spatially-extended TeV PWNe HESS~J1825--137, HESS~J1837--069, and HESS~J1841-055 that are 2FHL sources \cite{Ackermann:2015uya}.

For a soft Galactic spectrum, a plausible assumption could be that the lowest energy IceCube events are more relevant for comparing to TeV gamma-ray data.  Towards this end, we divide the IceCube data into two sets with roughly equal event numbers in each, with a cut at estimated $E_\nu \!=\! 100$~TeV.

The middle and bottom panels of Fig.~\ref{icetev} display the estimated IceCube densities for these sets constructed in the same manner as before.  To the $E_\nu \!=\! 20$--100~TeV figure, we show TeV density contours using a $10^\circ$ Gaussian.  Since 23 out of these 27 events are showers, this may be a more reasonable spread for what IceCube would see if associated.

For the $E_\nu \!>\! 100$~TeV set, we display the fields visible to the CASA-MIA, KASCADE, and GRAPES ground arrays yielding gamma-ray limits from this range in Fig.~\ref{SED}, with arrows pointing in the direction covered.  While the break of these limits below the IceCube flux seemingly disfavors a sizable Galactic contribution, Fig.~\ref{icetev} shows a region containing a fraction of these IceCube events is not yet directly limited by diffuse $\gtrsim\! 100$~TeV gamma-ray data, as discussed in \cite{Ahlers2014}.

\section{Is Sgr~A* a Pevatron Now?}
\label{peva}
Perhaps the most intriguing astrophysical interpretation of the TeV signal seen from the GC is the interaction of $>\,$PeV protons accelerated in the vicinity of the supermassive black hole (SMBH) associated with Sgr~A* (e.g., \cite{Aharonian2005}).  Accounting for recent advances in observations, we reexamine such a scenario.  Several novel features can in principle arise, including TeV synchrotron emission (well beyond the classical $\sim 100$~MeV radiation reaction limit), multi-PeV neutrinos, and Sgr~A* mm emission (e.g., \cite{Yuan2014}) causing an earlier cutoff in gamma rays due to $\gamma \gamma \rightarrow e^+ e^-$ attenuation.

We consider interactions within the inner accretion flow using properties from Sgr~A* models of \cite{Dexter2013} with some simplifying assumptions.  The emission spectra span from radio to IR (model 915h) and we take this to be spherical and extending it as $\epsilon_\gamma^{-2}$ for $\epsilon_\gamma \!\gtrsim\! 1$~eV.  We consider a generic smoothly-broken power law source proton spectrum
\begin{equation}
      \frac{dN}{dE_p}  =   f_\nu
       \left[\left(E/E_b\right)^{\alpha \eta} + \left(E/E_b\right)^{\beta \eta} \right]^{1/\eta}  e^{-E/E_c} \,,
\label{specfit}
\end{equation}
using $\alpha \!=\! 0$, $\beta \!=\! -2$, $E_1 \!=\! 1.5$~GeV, and  $E_c \!=\! 50$~PeV, with $L_p \!=\! 10^{38}\,$erg~s$^{-1}$ utilizing $\sim\!10\,$\% conversion of a $\sim\! {\rm few} \times 10^{-8} M_\odot\, {\rm yr}^{-1}$ transrelativistic SMBH accretion rate.

We assume all interactions to occur within a distance from the black hole of $3\, r_g$, with $r_g \!\simeq\! 6 \times 10^{11}$~cm, comparable to the IR emitting regions of the model, leading to the $\gamma \gamma$ attenuation curve shown in Fig.~3 of \cite{Kistler2015}.  We note that mm emission mostly arises from $\sim 5\, r_g$ and using a larger radius would decrease the compactness and thus the $p\gamma$ and $\gamma \gamma$ scattering rates.  For both regions $B \!\sim\! 50$~G, while the ambient proton density is $n_p \!\sim\! 10^5\,{\rm cm}^{-3}$ for the IR region and $\sim\! 10^7\,{\rm cm}^{-3}$ for the mm, we take $n_p \!\sim\! 10^6\,{\rm cm}^{-3}$.  We obtain spectra for secondary photons, electrons, and neutrinos from $pp$ scattering as in \cite{Kelner2006} and \cite{Kelner2008} for photopion ($p \gamma$;  finding comparable results using \cite{Hummer2010}).  The large ratio of magnetic to photon energy densities in these regions would result in rapid electron energy losses via synchrotron rather than IC, so we neglect a complete cascade treatment, using methods of \cite{Kistler2015} (note that for a couple orders of magnitude larger $B$ one should take into account QED; \cite{Schwinger1949}) for both secondary electrons arising from pion decays and $\gamma \gamma \rightarrow e^+ e^-$ within the source region.

\begin{figure}[t!]
\hspace*{-0.3cm}
\includegraphics[width=1.05\columnwidth,clip=true]{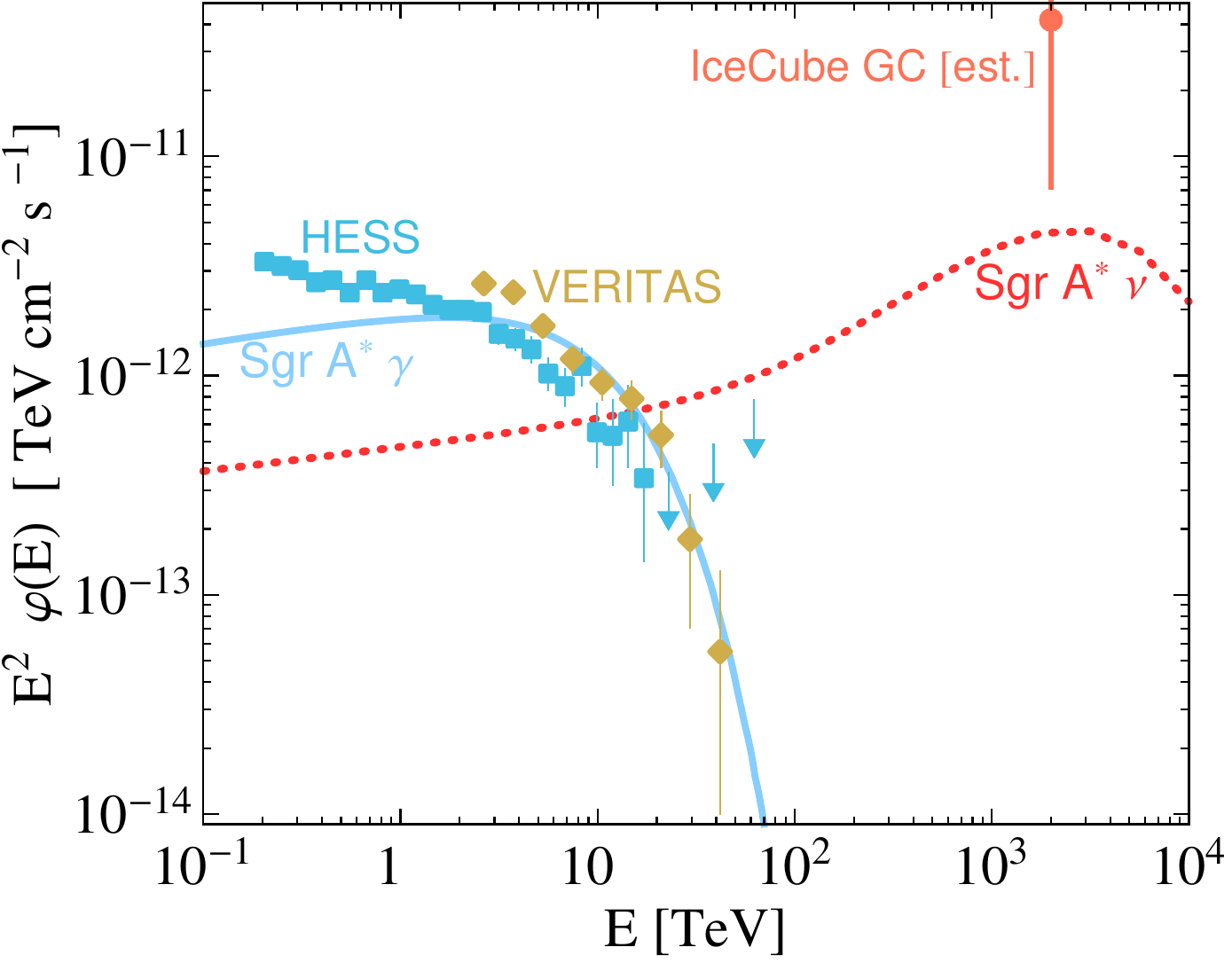}
\caption{Gamma-ray ({\it solid line}) and neutrino spectra ({\it dotted line}) from Sgr~A* $pp$ and $p \gamma$ interactions, where luminosities are scaled to reach the $\sim\! 1\,$TeV data, a factor of $7 \!\times\! 10^4$ above a scenario that assumes free escape from the inner flow within $t \!\sim\! 3\,r_g/c$.  GC source TeV gamma-ray data are from HESS \cite{Viana2013} and VERITAS \cite{Smith2015}.
\label{pevfig}}
\end{figure}

Fig.~\ref{pevfig} shows the resulting gamma-ray and neutrino spectra with the normalizations set to match the $\sim\! 1$~TeV data.  This corresponds to an increase by  a factor of $7 \!\times\! 10^4$ over the assumption that protons escape the inner region within $t \!\sim\! 3\,r_g/c$.  Doing so, we can get neutrino emission peaking at $\sim\,$few~PeV, consistent with the $\gtrsim\,$1~PeV IceCube shower event centered within $\sim\! 1^\circ$ of Sgr~A* in Fig.~\ref{iceall}, although with angular resolution of $\sim\! 10^\circ$ \cite{Aartsen2013,Aartsen2013b}, if some luck is also involved (we have estimated the implied flux in Fig.~\ref{pevfig}).

Why are the free escape fluxes so low?  Simply, the interaction rate for the above parameters is low.  We consider ways to boost this, although we should be wary of breaking something.  First, the mm-IR spectrum of Sgr~A* is softer than $\epsilon_\gamma^{-2}$, implying a far larger number density of target photons for $p \gamma$ at mm energies ($\epsilon_\gamma \!\sim\! 10^{-3}$~eV), for which the threshold proton energy for photopion production is proportionately higher ($E_p \!\gtrsim\! 1000$~PeV).  Since the pion spectrum produced via this process is thus rather hard ($dN/dE_\pi \sim E_\pi^{-1}$), the last decade of the proton spectrum is most relevant for energy extraction via $p \gamma$.  Indeed, \cite{Aharonian2005} considered a few Sgr~A* scenarios, favoring $p \gamma$ with  $E_p^{\rm max} \!\gtrsim\! 10^{18}$~eV.

The lack of higher-energy neutrino events in IceCube from the GC thus far in principle does allow one to extract more energy from the $p \gamma$ process by extending this hard secondary spectrum to higher energies.  However, consideration of the basic criterion of $E_p^{\rm max} \sim e B R$ requires a field of $\sim\! 10^4$~G for $R \!\lesssim\! 10\,r_g$.  This would tend to lead to a harder gamma-ray spectrum, though, since reprocessing of $\gamma \gamma \rightarrow e^+ e^-$ secondaries via synchrotron becomes more dominant.  It was simultaneously assumed in \cite{Aharonian2005} that $B\sim 10$~G to allow for IC cascades (and a somewhat larger target background).

One could invoke a source proton spectrum harder than $E_p^{-2}$, which for the same total $L_p$ would put more energy into the last decade, decreasing the boost needed to reach the TeV data by $\sim 10$ for $E_p^{-1}$ (e.g., \cite{Aharonian2005} only considered $L_p$ as near $E_p^{\rm max} \!\sim\! 10^{18}$~eV), at the cost of requiring an explanation for such a spectrum and, again, a harder gamma-ray spectrum than displayed in Fig.~\ref{pevfig}.

We can enhance the 1--10~TeV flux by increasing $n_p$, although this would be more necessary for a lower $E_p^{\rm max}$, since $p \gamma$ interactions would decrease in relative importance.  This would also increase the production of lower-energy secondary electrons and thus synchrotron emission extending down to lower energies, particularly in the {\it Chandra} range for $E_e \!\lesssim\! 100$~GeV.  Of a total $2\!-\!10$~keV flux from Sgr~A* of $\sim\! 3 \!\times\! 10^{33}\,$erg~s$^{-1}$, \cite{Wang2013} estimates that $\sim\! 20$\% is from the inner accretion flow and \cite{Neilsen2015} derives a $\sim\! 10$\% fraction as arising from faint Sgr~A* flares.

The model in Fig.~\ref{pevfig} is nominally consistent with these, although we also have not considered primary $e^-$ acceleration, for which there is evidence from the hard X-ray flares detected by {\it NuSTAR} \cite{Barriere2014}, tacitly assuming that these are due to some other mechanism or variability in the $p/e$ efficiency.  While a softer proton spectrum could better account for $< \,$TeV gamma rays, it would be at the cost of a larger boost, fewer $p \gamma$ interactions, and an increase in synchrotron to above the {\it Chandra} flux.  Similarly, a lower $E_p^{\rm max}$ would decrease the $p \gamma$ fluxes, requiring a greater $pp$ contribution leading to the same issues.

If the magnetic field structure near the SMBH is mostly toroidal, then diffusion that needs to cross field lines could inhibit particle diffusion outwards, although it appears difficult to get more than a factor of $\sim 100$ via such a delay.  Such a scenario is also complicated by the proximity of the SMBH, which could spell doom for such lingerers.  However, a long containment time would lengthen timescales associated with the TeV emission, consistent with the lack of variability or increases during X-ray flares in HESS data \cite{Aharonian2009}.  We also note that for $B \!\sim\! 50$~G and assuming the electron cooling break is beyond the end of the unbroken {\it NuSTAR} spectra could suggest proton acceleration to $\gtrsim\!10$~PeV if the entire $\sim\,$few$\,\times$$\,1000\,$s flare durations are available, although most energy would need to go into protons to be relevant to TeV data.

We could consider proton interactions beyond the immediate vicinity of the SMBH; however, the benefit of the enhanced pair opacity beyond the GC curve in Fig.~3 of \cite{Kistler2015} is then no longer available, so that the cutoff in the TeV data would have to be due to a turndown in the proton spectrum at  $E_p \!\lesssim\! 100$~TeV.  The neutrino fluxes would be similar to those previously considered (e.g., \cite{Crocker2005,Kistler2006}).  The $\gtrsim \!10$~TeV attenuation can be somewhat increased if $pp$ interactions exclusively occur in strongly IR emitting regions, perhaps by a factor of a few.  However, the IR and FIR maps do not well coincide and most of the gas mass available for $pp$ scattering is in the colder CND.  Other types of model also claim gamma rays from near Sgr~A* or cosmic rays in its surroundings (e.g., \cite{Atoyan2004,Quataert2005,Liu2006,Ballantyne2007,Ballantyne2011,Linden2012,Kusunose2012,YusefZadeh2013,Guo:2013yva,Fujita2015,Wang2006,Hinton2007,Mori2015,Kistler2015b}).  A gamma-ray spectral break at $\sim\! 10$~TeV may also simply reflect a break in the present proton spectrum due to diffusive escape above $\sim\! 10^{14}\,$eV due to similar physics as may result in the PeV ``knee'' in the cosmic-ray spectrum measured at Earth.

In total, it appears difficult to account for the TeV data using Sgr~A* without a large persistent reservoir of energetic protons within a few gravitational radii of the SMBH.  A lack of hadronic production from Sgr~A* may actually be considered beneficial, since it may indicate that such sources elsewhere in the universe are optically thin to escape, as inferred from relating cosmic-ray data to IceCube (e.g., \cite{Kistler2014}).    Also, the central $\sim\,$100~pc has the highest SN rate density in the Milky Way \cite{Ponti2015,OLeary2015}, so acceleration by the SMBH is not necessarily required for a harder, enhanced cosmic-ray density, likely also contributing to the TeV flux near the GC.

\section{Discussion and Conclusions}
\label{concl}
The presence of cosmic-ray protons up to $E_p \!\gtrsim\! 1$~PeV and TeV gamma-ray sources imply that a Galactic TeV neutrino flux should be present at some level.
The population of Galactic TeV gamma-ray sources generally appear to be lacking at $E_\nu \!\gtrsim\! 10$~TeV (e.g., \cite{Kistler2006,Kappes:2006fg}), even if the emission is hadronic.  Diffuse TeV gamma-ray measurements are concentrated along the Galactic disk and relatively low in flux \cite{Aharonian:2005kn,Abramowski:2014sla,Deil2015,Abdo:2008if,Bartoli:2015era,Abeysekara:2015qba}, consistent with our simple counting exercise.  While more sophisticated tests are available (e.g., \cite{Ahlers:2015moa,Neronov:2015osa}), and we have looked at various distributions of these data, the figures presented adequately illustrate the situation.

The concentration of {\it Fermi} TeV gamma rays in the inner galaxy suggests $|l| \!\lesssim\! 30^\circ$ is of greatest interest for improvement.  The HESS Galactic plane survey \cite{Aharonian:2005kn,Abramowski:2014sla,Deil2015} and HAWC \cite{Abeysekara:2015qba} are operating in this region presently, although very high energies are not well covered, since the limits shown in Fig.~\ref{SED} are all from detectors in the northern hemisphere.  IceCube has some sensitivity to PeV gamma rays \cite{Aartsen:2012gka}, and along with a larger detector \cite{Aartsen:2014njl} one would like to have absolute gamma-ray constraints from the downgoing region.
It seems most likely that very-high energy Galactic emission would also be constrained to a narrow region around the plane; however, it would also be useful if archival data is used to place limits considering larger regions around the Galactic plane.

Electromagnetic shower data from HESS and VERITAS in Fig.~\ref{SED} suggest that the flux at $\sim\! 1 \!-\! 10$~TeV is also low compared to the sky-averaged IceCube flux in regions off the plane, while HAWC plans to significantly improve on their limit \cite{Pretz2015}.  Isotropic TeV electromagnetic cascades can also be ascribed to $e^\pm$ from pulsars (see, e.g., \cite{Kistler2012,Linden:2013mqa,Venter:2015gga}), so it will be important to determine this contribution, as CALET \cite{Adriani:2015pqa} and CREST \cite{Musser2015} aim to accomplish.  The low {\it Fermi} TeV gamma-ray flux off the Galactic plane implies that the $e^- \!+\! e^+$ data in this range are fairly free of gamma-ray emission.

{\it Fermi} has demonstrated a good ability to distinguish charged particles from gamma rays \cite{Abdo:2009zk}.  Although instrument response function sensitivities drop off at $\gtrsim$~TeV presently \cite{Atwood:2013rka,LATpage}, if a 10~TeV gamma ray hits the instrument it will definitely result in a signal, only a very complex one.  As compared earlier, the neutrino effective area at $\sim\! 10$~TeV is $\sim\! 2~{\rm m}^2$ for the all-sky search in \cite{Niederhausen2015}, comparable to the LAT effective instrument area estimated from the $\sim\! 2.5\,{\rm m}^2\,$sr acceptance below $1$~TeV \cite{Atwood:2013rka}, so in principle can be statistically comparable for the forthcoming period.  From Fig.~\ref{SED}, several hundred multi-TeV $\gamma$ events could have struck over $\sim\! 7$~yr of operation.  Even if energy resolution is limited in saturating events, it may at least be possible to better constrain the Galactic contribution than done here.

We have also examined spectra from dark matter decays, which has generated some interest mostly at higher energies (due to the early clustering of neutrino events $\sim\! 1$~PeV, e.g., \cite{Feldstein:2013kka,Rott:2014kfa,Esmaili:2014rma,Murase2015}).  Neutrinos at $\gtrsim\! 10$~TeV can result for various masses and decay channels \cite{Cirelli2011}, with more or less gamma rays generally possible.
In order to have a noticeable effect on the IceCube flux an associated gamma-ray flux may be detectable.  It is straightforward to integrate over a dark matter profile to find the decay flux profile (e.g., \cite{Yuksel:2007dr}), with $\sim\! 1/2$ of the total arising from within $5 \!-\! 60^\circ$ of the GC (not including a redshifted extragalactic component of lesser importance).  Annihilation neutrinos would be even more centrally concentrated (e.g., \cite{Yuksel:2007ac}).

Cursory examination of off-plane regions near the GC in Fig.~\ref{fermitev} shows no obvious declining, symmetric signal as would be expected (see also \cite{Massari:2015xea}), although with {\it Fermi} one is attempting to catch $>\! 30$~TeV dark matter by the tail.  The {\it Fermi} bubbles \cite{Su2010,Fermi-LAT:2014sfa} are also of interest for neutrinos (e.g., \cite{Crocker:2010dg,Lunardini:2015laa,Lacki2014}).  The upper limits obtained at high-energy \cite{Fermi-LAT:2014sfa} are also consistent with a low number of counts in Fig.~\ref{fermitev}.

It would be of great benefit to measure and map the Galactic neutrino emission, not only for understanding hadronic processes in the Milky Way, but also as a test of difficult sources such as the one at the Galactic Center that may be due to Sgr~A* or neutrino-free PWN emission \cite{Wang2006,Hinton2007,Mori2015,Kistler2015b}.  This determination will also permit better understanding of the extragalactic component of the IceCube flux and understanding of their sources (e.g., also \cite{Waxman:1998yy,Mannheim2001,Laha2013,Baerwald:2013pu,Anchordoqui:2013qsi,Chen:2014gxa,Padovani:2015mba,Kistler2015c,Murase:2015ndr}).\\

%
We are quite grateful to Matthew Wood for assistance with {\it Fermi} tools, Hasan Yuksel for computational assistance, Jason Dexter for Sgr~A* discussions,
and James Lang for comments.
We also thank the organizers of INT Program INT-15-2a ``Neutrino Astrophysics and Fundamental Properties'' for hospitality during part of this project and
the {\it Fermi} LAT Collaboration for the availability of data and {\it Fermi} Science Tools.
MDK acknowledges support provided by Department of Energy contract DE-AC02-76SF00515, and the KIPAC Kavli Fellowship made possible by The Kavli Foundation.



\appendix*

\onecolumngrid

\section{Supplemental Figures}
\label{app1}

We include below additional density map figures of IceCube events without further adornment.

\begin{figure*}[t!]
\vspace*{-0.85cm}
\includegraphics[width=0.9 \columnwidth,clip=true]{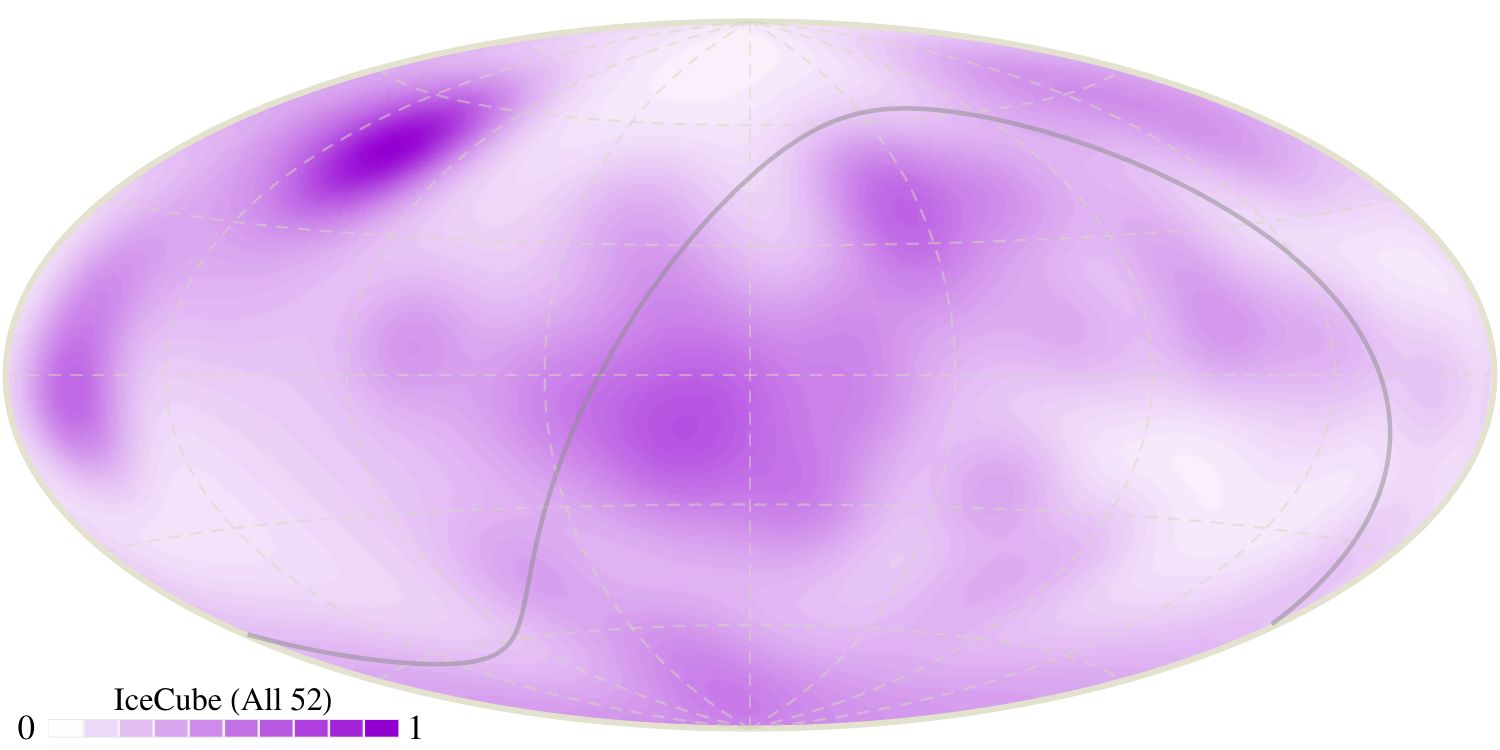}\\
\vspace*{-0.1cm}
\includegraphics[width=0.9 \columnwidth,clip=true]{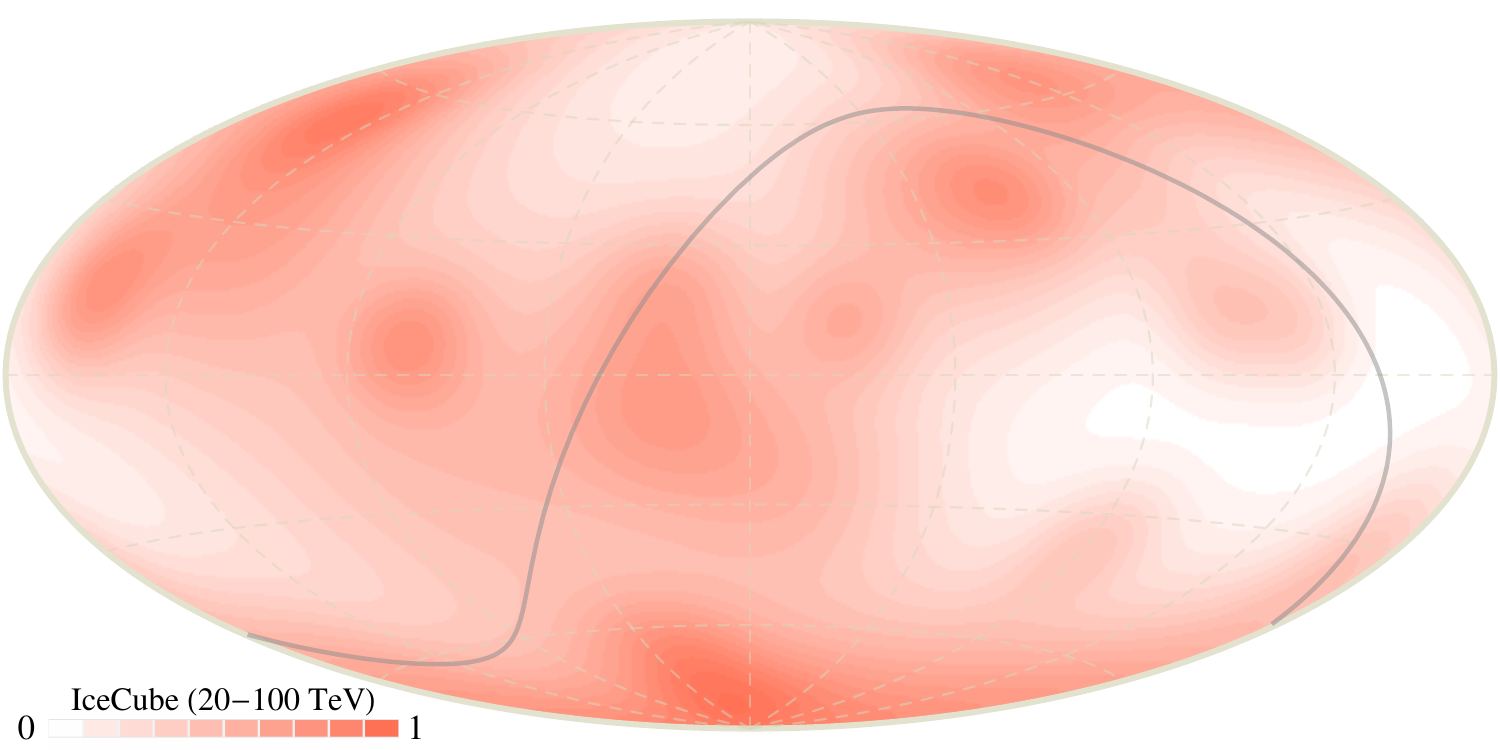}\\
\vspace*{-0.1cm}
\includegraphics[width=0.9 \columnwidth,clip=true]{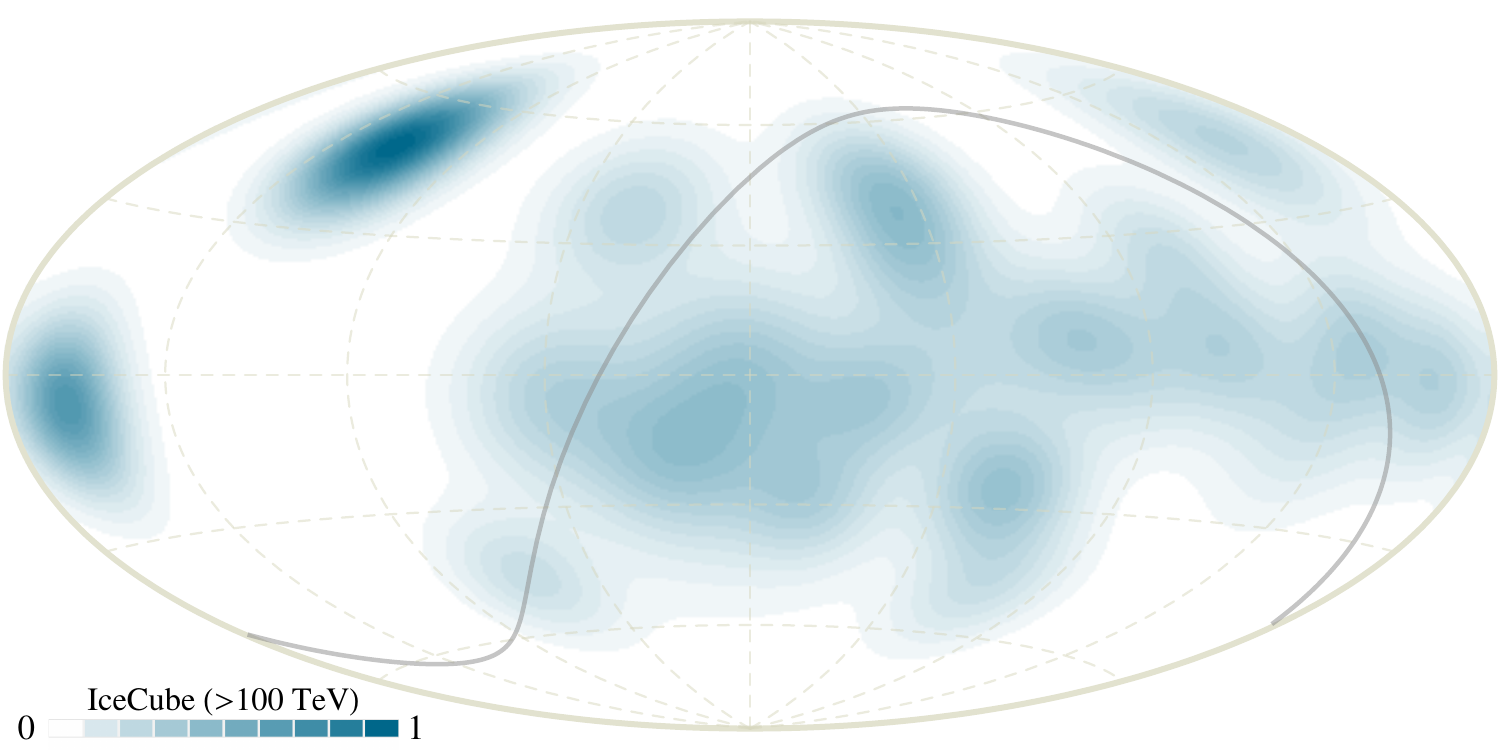}
\vspace*{-0.2cm}
\caption{Estimated density distributions of all IceCube neutrino events ({\it top}) and with estimated neutrino energy ranges as in Fig.~\ref{icetev}.
\label{icetev3}}
\end{figure*}

\begin{figure*}[t!]
\vspace*{-0.85cm}
\includegraphics[width=0.9 \columnwidth,clip=true]{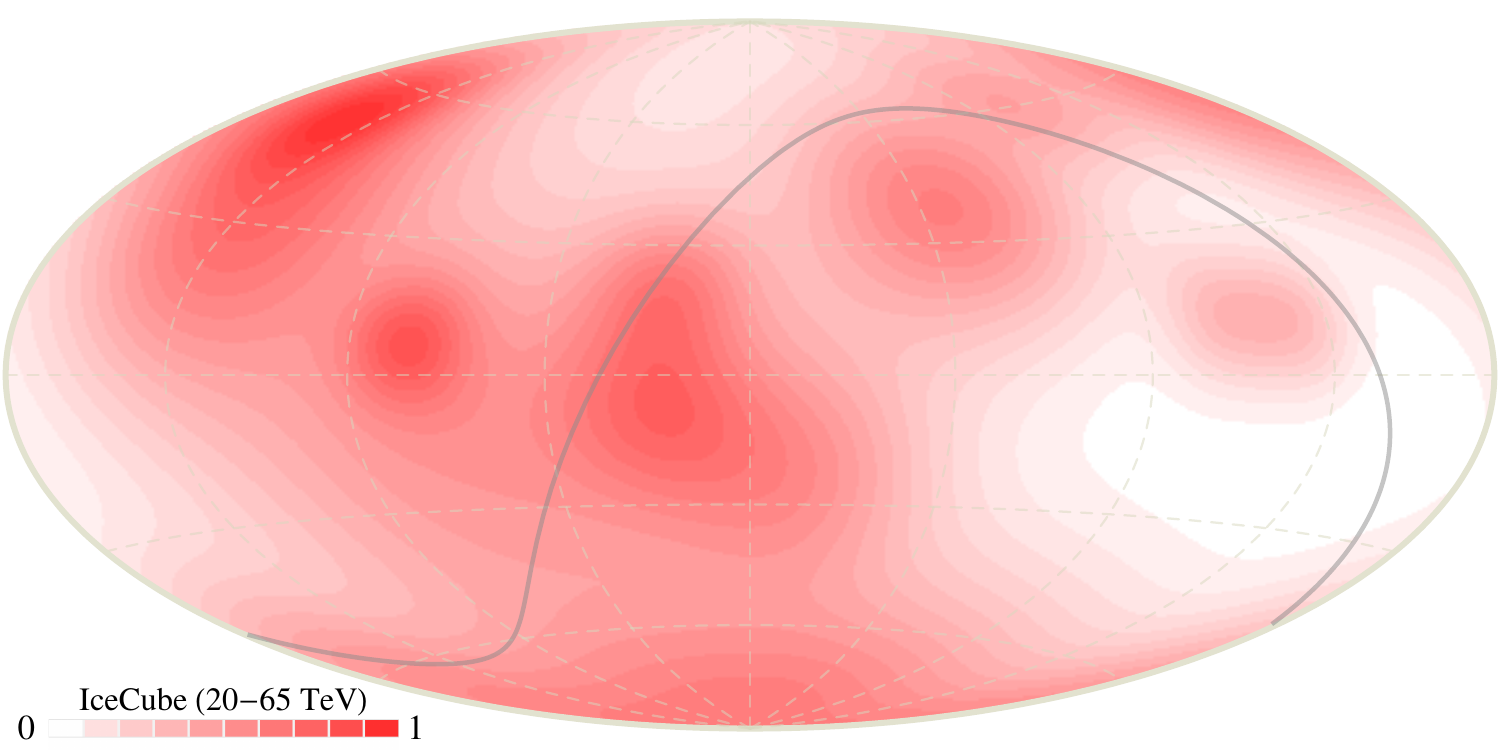}\\
\vspace*{-0.1cm}
\includegraphics[width=0.9 \columnwidth,clip=true]{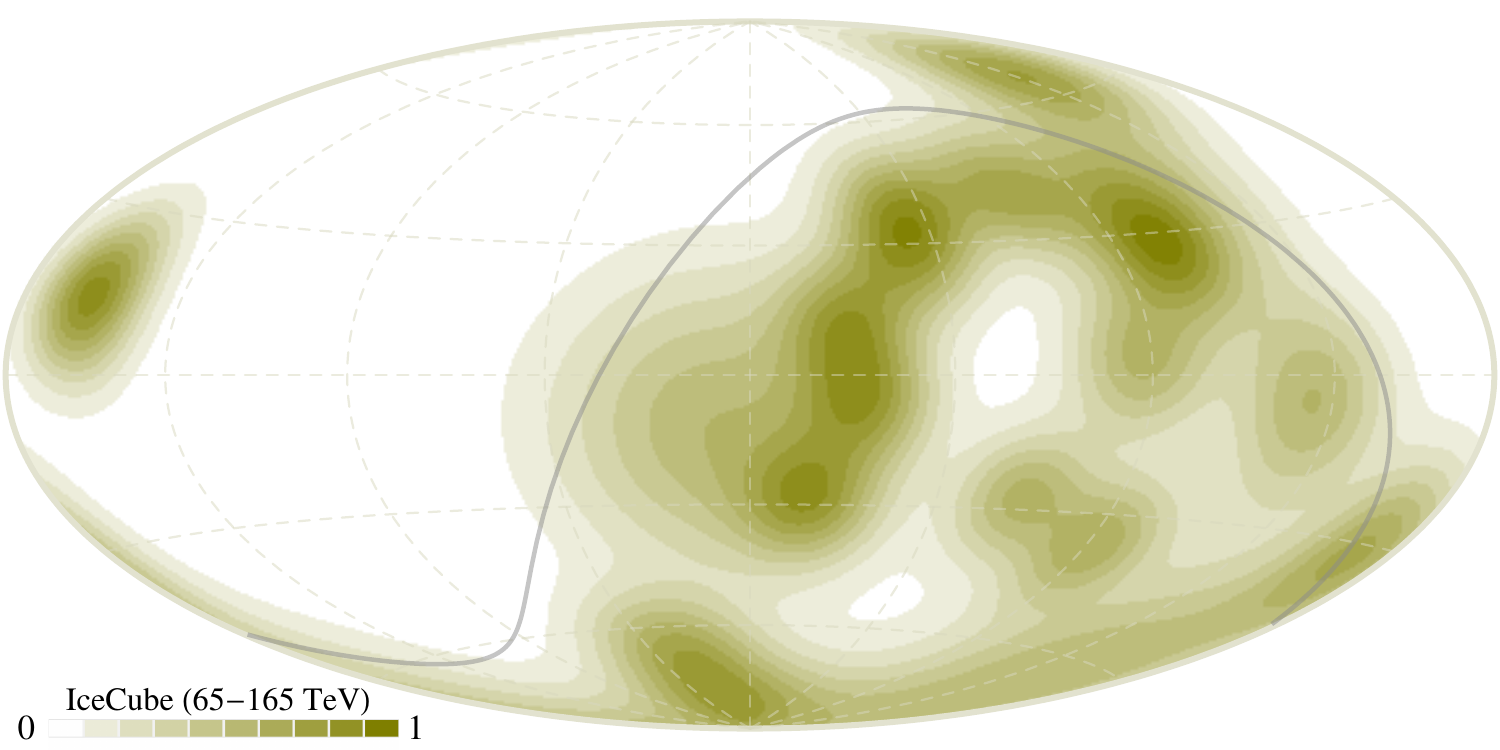}\\
\vspace*{-0.1cm}
\includegraphics[width=0.9 \columnwidth,clip=true]{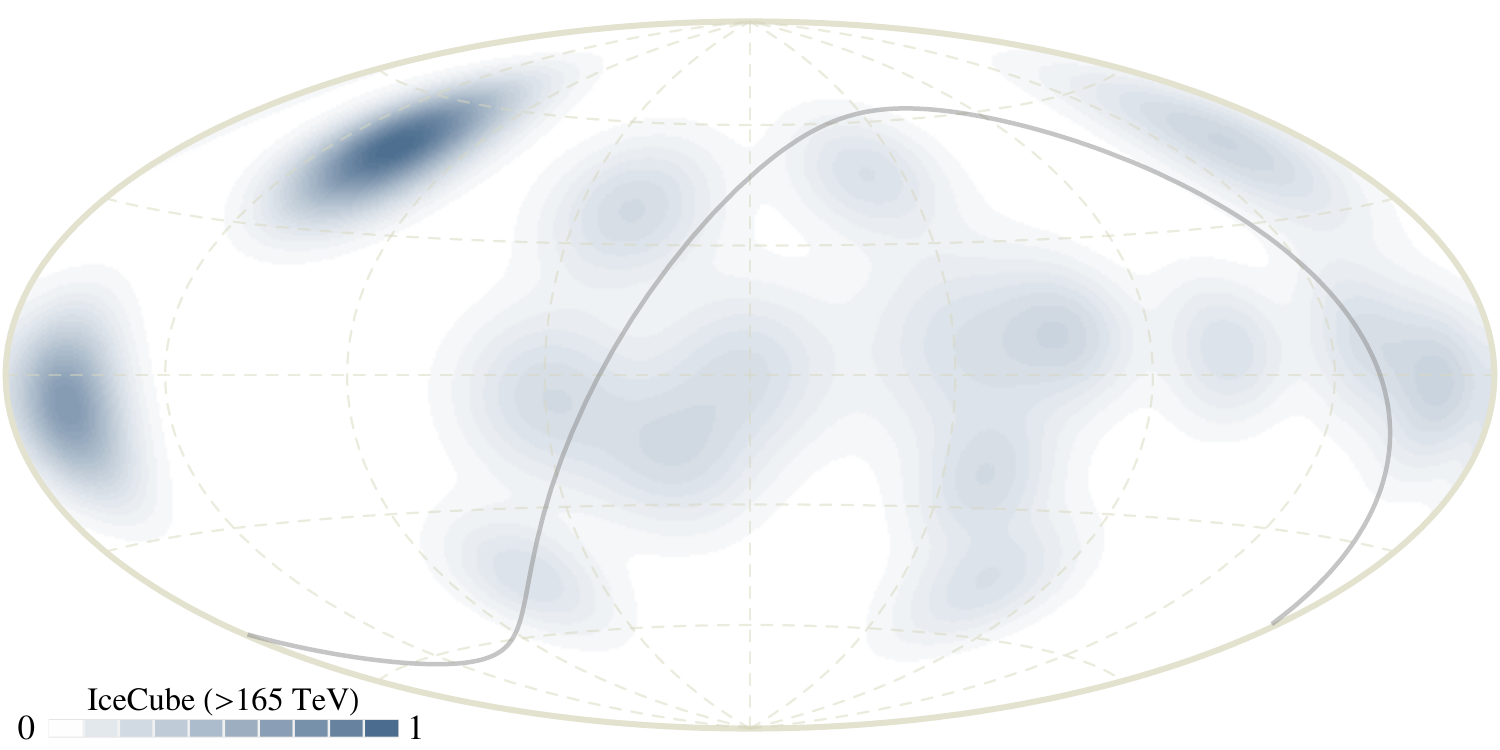}
\vspace*{-0.2cm}
\caption{Estimated density distribution of IceCube neutrino events broken down into three estimated neutrino energy ranges: 20--65~TeV ({\it top}; 17 events), 65--165~TeV ({\it middle}; 19 events), and 165--2000~TeV ({\it bottom}; 16 events).
\label{icetev4}}
\end{figure*}

\end{document}